\documentclass[doublecol]{epl2}

\usepackage{graphicx}
\usepackage{dcolumn}
\usepackage{bm}
\usepackage{mathrsfs}
\usepackage{amssymb}

\title{Monte Carlo simulations of the randomly forced Burgers equation}%

\author{P. D\"{u}ben\inst{1} \and D. Homeier\inst{1}\thanks{E-Mail: \email{homeierd@uni-muenster.de}} \and K. Jansen\inst{2} \and D. Mesterhazy\inst{3} \and G. M\"{u}nster\inst{1} \and C. Urbach\inst{3}}
\shortauthor{D\"{u}ben, Homeier, Jansen, Mesterhazy, M\"{u}nster and Urbach}
\institute{
  \inst{1} Westf\"{a}lische Wilhelms--Universit\"{a}t M\"{u}nster\\
  \inst{2} DESY, Zeuthen\\
  \inst{3} Humboldt--Universit\"{a}t zu Berlin
}

\pacs{02.70.Uu}{Applications of Monte Carlo methods}
\pacs{05.20.Jj}{Statistical mechanics of classical fluids}
\pacs{47.40.Nm}{Shock wave interactions and shock effects}

\abstract{
The behaviour of the one--dimensional random--forced Burgers equation is
investigated in the path integral formalism, using a discrete
space--time lattice. We show that by means of Monte Carlo methods one
may evaluate observables, such as structure functions, as ensemble
averages over different field realizations. The regularization of shock
solutions to the zero--viscosity limit (Hopf-eq.) eventually leads to
constraints on lattice parameters, required for the stability of the
simulations. Insight into the formation of localized structures
(shocks) and their dynamics is obtained.}

\begin{document}
\maketitle

\section{I. Introduction}
\noindent
The random--force--driven Burgers equation
\begin{equation}
\partial_t u + u \partial_x u - \nu \partial_x^2 u = f
\label{dh:eqburg}
\end{equation}
may be interpreted as a model for compressible hydrodynamic turbulence,
describing acoustic perturbations in the reference frame moving with the
velocity of sound \cite{dh:bibburgers, dh:bibfalkovich}. Here $u$ is
the velocity, $f$ the random forcing, and $\nu$ is the kinematic
viscosity. One generally assumes the force to be white--in--time,
displaying Gaussian statistics
\begin{equation}
\langle\, f(x,t) f(x',t') \,\rangle=\chi(x-x')\,\delta(t-t')
\end{equation}
so that the properties of the external pumping are completely
characterized by the covariance $\chi$. The turbulent state is
maintained by a large--scale external force with correlation length $L$.
Finite $\nu > 0$ and energy dissipation $\epsilon = \chi(0)$ provide a
dissipation scale \mbox{$\eta = (\nu^3 / \epsilon)^{1/4} > 0$}. The
dissipation is related to the Reynolds number Re by the definition
\mbox{$\epsilon\equiv \textrm{Re}^3 \nu^3/L^4$}. Thus the
characteristic velocity is $u_L=(\epsilon L)^{1/3} $. In the limit of
large Reynolds number \mbox{$\mathrm{Re}= ( L/\eta )^{4/3} \gg 1$} these
scales separate, so one expects a turbulent cascade. Evaluating moments
of Galilean invariant velocity differences \mbox{$w(r)\equiv
u(r)-u(0)$}, i.e.~structure functions
\begin{equation}
S_p(r)\equiv \big\langle [w(r)]^p\big\rangle, \label{dh:struc}
\end{equation}
$p>0$, where $r$ is the displacement, one therefore expects a scaling
behavior in the inertial interval $L \gg r \gg \eta$. From dimensional
considerations we find $S_p(r) \propto (\epsilon r )^{\zeta_p}$, and
scale--invariance implies $\zeta_p=p/3$. However, the formation of
shocks in Burgers turbulence leads to a strong intermittency and a
bifractal scaling of the form $\zeta_p=\min(p,1)$, see
e.g.~\cite{dh:bibbec}.

We would like to emphasize that the average $\langle\,\cdots\,\rangle$
in Eq.~(\ref{dh:struc}) and in the following is usually considered to be
an average over all times at a given place, or over all places at a
given time; linked by the Taylor--hypothesis in statistically homogenous
turbulence. In this article these averages are represented as ensemble
averages, as discussed below.

Numerical investigations of hydrodynamic systems are commonly performed
by means of direct simulations or variants thereof which amounts to
integrating the equations of motion. In this article we follow a
completely different approach, namely to evaluate the corresponding path
integrals by Monte Carlo simulations. The clear physical picture of
Burgers equation, and the fact that the intermittent structures are
well--known -- they correspond to shocks with a large negative velocity
gradient -- makes the Burgers equation an attractive setting to study
Monte Carlo methods in turbulence theory. Further, a number of
technical advantages leads to the unique possibility of investigating
intermittent structures in turbulent flow. In particular, Burgers
equation is local while incompressibility acts as a nonlocal interaction
in Navier--Stokes turbulence which eventually complicates issues due the
presence of the functional determinant arising in the path integral
formulation of Eq.~(\ref{dh:eqburg}) (see Sec.~II). Also the
dissipation scale $\eta$ provides a UV--regularization of shock
structures. Finally, a huge variety of analytical methods has been
applied to Burgers equation, giving results that can directly be
compared to numerical calculations. However, Burgers equation is
certainly also interesting on its own (e.g.~in cosmology), serving as
model for systems with an interplay of random processes and coherent
structures; for an overview see \cite{dh:bibbec}.

It is the aim of this paper to demonstrate the feasibility of Monte
Carlo simulations in the path integral formalism. In particular we will
show that respecting certain constraints on the parameters of the model,
stable simulations can be performed. We will demonstrate with an
example that this way structure functions can be computed precisely.

This paper proceeds as follows. In Sec.~II we shortly introduce the
path integral formulation for Burgers equation. Then in Sec.~III the
numerical scheme is discussed in detail and in Sec.~IV the numerical
results are given.

\section{II. Path Integral for Burgers equation}
\noindent
The path integral for the randomly forced Burgers equation is introduced
via the Martin--Siggia--Rose formalism \cite{dh:bibmsr} where one has
\begin{equation}
Z = \int\!Du\,Dp\,\exp \big(- S[u,p] \big)
\label{MSRg}
\end{equation}
with the conjugate momenta $p$, and the action is \mbox{$S[u,p]
=\int\!dt\,\mathcal{L}[u,p]$}. The Lagrangian is given by
\begin{equation}
\mathcal{L} = - i \int\!dx\, p\,\big(\partial_t u + u\partial_x u - \nu
\partial^2_x u \big) + \frac{1}{2}\int\!dx\,p\,( \chi\ast p )\,,
\label{MSRl}
\end{equation}
where $\ast$ denotes the convolution.
Performing the Gaussian integration over the conjugate field, one
arrives at
\begin{equation}
\mathcal{L} = \frac{1}{2}\int\!dx\, N[u]\,\big( \chi^{-1} \ast N[u]\big)
\label{numl}
\end{equation}
where $N[u]\equiv \partial_t u + u\partial_x u - \nu \partial^2_x u$.
The functional determinant, arising in principle in the derivation of
(\ref{MSRg}), (\ref{MSRl}) does not contribute to the forced Burgers
equation due to the local interaction (see e.g.~\cite{dh:bibhmpv}).
Thus the integration measure is properly defined, and the integration is over 
all fields $u$ (and $p$) in one dimension.
In contrast, in the case of the Navier--Stokes equation, where the effect of
pressure is not negligible, the corresponding non--local interaction leads
to a non--vanishing functional determinant that has to be treated using the 
Fadeev--Popov procedure \cite{dh:mydiss}.

As (\ref{MSRg}), (\ref{MSRl}) are Galilean invariant, the computation of
non-invariant quantities such as n--point correlation functions would require a
gauge fixing \cite{dh:bh2007}. However, since we are interested in evaluating
structure functions, i.e.\ manifestly Galilean-invariant quantities, this procedure
does not have to be performed.

Then the expectation values of Galilean invariant observables $O[u(x)]$ are 
calculated as ensemble-averages
\begin{equation}
\big\langle O[u] \big\rangle
= \frac{\int\!Du\, O[u]\, \exp \big(- S[u] \big)}
{\int\!Du\,\exp \big(- S[u] \big)}\,.
\label{nptf}
\end{equation}
Extending the time integration in the action to infinity renders the
averages stationary (see e.g.~\cite{dh:g1984}).

Using the path integral (\ref{MSRg}),(\ref{MSRl}) it has been shown that
intermittent statistics of Burgers equation can be understood in terms
of instanton solutions \cite{dh:bibfklm,dh:bibbfkl,dh:bibgm}. These
solutions break Galilean invariance which has been shown to account for
intermittency \cite{dh:bibpolyakov}.

\section{III. Monte Carlo Simulations}
\noindent
Using the path integral formalism we calculate ensemble averages like
(\ref{nptf}) numerically,
where the Lagrangian is given by (\ref{numl}). Defining the theory on a
finite size lattice, the integration is performed over grid variables
$u_i(n)$, i.e.~the field $u$ defined at lattice sites $(i,n)$ where $i$
denotes the discretized space, and $n$ discretized time points. Then we
obtain
\begin{equation}
\langle O \rangle = \int \prod_{i,n} d u_i(n)\,\,O[\{u_i(n)\}]\,
\exp \big(- S[\{u_i(n)\}]\big)
\label{dPI}
\end{equation}
for the observable computed in the discretized theory. Here the
discretized action $S[\{u_i(n)\}]$ may be so chosen that the single-site
action $S\big(u_i(n)\big)$ is quadratic:
\begin{equation}
S\big(u_i(n)\big)= A_i(n) \big[u_i(n)-C_i(n)\big]^2 + B_i(n)~,
\label{action}
\end{equation}
where $A_i(n)>0$. The terms $A_i(n)$, $C_i(n)$ and $B_i(n)$ will in
general depend on the field at different nodes on the lattice. The
action is locally minimized by choosing $u_i(n)\simeq C_i(n)$ thus
giving the main contribution to the integral in (\ref{dPI}). This way
one may evaluate the whole integral, by successively integrating over
different lattice sites. Successive application of such sweeps, will
then generate the Markov chain $\{u^{(k)}_i(n)\}$ of configurations in
the $k$--th step of the process. The quadratic form of the action in
(\ref{action}) allows to apply a heat-bath Monte Carlo algorithm to
update single field variables on the lattice. However, it should be
emphasized that writing the single--site action as (\ref{action})
imposes a restriction on the possible form of the nonlinear term $u
\partial_x u$. In its discretized version, it has to be at most linear
in $u_i(n)$. Thus we are led to define the nonlinear term as \mbox{$u
\partial_x u \rightarrow u_i(n) u_i'(n)$}, where the derivative is
written symmetrically with the simplest choice being \mbox{$u_i'(n) =
\big(u_{i+1}(n) - u_{i-1}(n)\big)/2 \Delta x$}. The time derivative and
the viscous term are discretized in the usual way.

The above action was put onto a rectangular lattice with
$N_{\textrm{x}}$ sites in space-- and $N_{\textrm{t}}$ sites in
time--direction. Lattice spacings are denoted correspondingly by
$\Delta x$ and $\Delta t$. Monte Carlo simulations were performed for
different lattice sizes, ranging from
e.g.~$(N_{\textrm{x}}=16)\times(N_{\textrm{t}}=4)$, to larger lattices
of up to approximately $(N_{\textrm{x}}=362)\times(N_{\textrm{t}}=4096)$
at fixed viscosity respectively. The Reynolds number ranged from
$\textrm{Re} = 2$ to $\textrm{Re} = 16$.

\subsection{Lattice parameters and Continuum Limit}
\noindent
In our simulations we chose the forcing characteristics as
\begin{equation}
\chi(x-x') = \chi(0)\,\exp\big(- |x-x'| / L \big)~,
\end{equation}
in agreement with the considerations in \cite{dh:bibbfkl}. In
Fourier--space this represents a noise $\tilde{\chi}(k)$ with an
IR-cutoff at wave-number $k \simeq L^{-1}$ and a power--law behavior for
$k\gg L^{-1}$, of the form $\chi(0)\, L^{1+\beta}\,k^{\beta}$, where
$\beta=-2$. That is a random force acting predominantly at large scales
$ \gtrsim \nolinebreak L$, leading to shock--formation at random positions in our
system \cite{dh:bibbec}.

To identify the lattice parameters with the corresponding constants of
the continuum theory, first notice that the viscosity has to be defined
as
\begin{equation}
\nu\equiv \tilde{\nu}\,(\Delta x )^2 / \Delta t
\end{equation}
where the arbitrary constant $\tilde{\nu}$ has been introduced, relating
units for space-- and time--measurements on the lattice. This can be
shown by performing the continuum limit in the case of the symmetric
random walk, leading to the diffusion equation. The so--defined $\nu$
gives us the Reynolds--number $\textrm{Re}$ according to the earlier
given relations, after $\chi(0)$ and $L$ have been specified.
The continuum limit is then performed by $\Delta x$, $\Delta t
\rightarrow 0$ while keeping $L$, $\nu$ and $\textrm{Re}$ constant.

In both space-- and time--direction periodic boundary conditions were
specified. Then one may expect that in the continuum limit the correct
physical behavior of the fields is assumed and correlation effects
induced by the boundary conditions are negligible.

Stability considerations lead to further constraints for the lattice
size, as will be explained in the following subsection.

\subsection{Stability and Regularization}
\noindent
As one would expect, the stability of the simulations over a large
number of Monte Carlo steps is a big issue. Indeed, if certain
restrictions on the lattice parameters are not taken care of,
the simulation terminates sooner or later due to instabilities.

The occurence of instabilities in our Monte Carlo simulations is related
to the existence of shock--like solutions of Eq.~(\ref{dh:eqburg}). To
obtain stable simulations, the shock structures have to be resolved on
the lattice, otherwise the action near the shock positions is
represented badly and the simulation tends to develop instabilities,
accumulating the overall energy of the configurations near the shocks.

In the limit of small viscosities it would be possible to regularize the
shocks by adding higher derivative terms or nonlinear terms to the
differential equation. Similar to the case of the Korteweg-de Vries
equation, where the term proportional to $\partial_x^3 u$ serves to
regularize shock waves \cite{dh:bibll,dh:bibv}, the Burgers equation can
be regularized by adding e.g.\ terms proportional to $\partial_t
\partial_x^2 u$ and $u \partial_x^3 u$ to it \cite{dh:bibbf}. This
would smear out the shocks and enable their proper treatment in the
simulation. As indicated by the shock solution of the unforced Burgers
equation
\begin{equation}
u = - A \tanh \frac{A}{2\nu} x\,,
\end{equation}
the viscosity term itself provides a valid regularization of shocks.
Therefore we decided to smooth out shocks by chosing the viscosity
appropriately. This amounts to resolve the dissipation scale $\eta$ on
the lattice. In terms of the Reynolds number this translates into the
constraint
\begin{equation}
\mathrm{Re} \lesssim L N_{\textrm{x}}\,.
\end{equation}
At given Reynolds number $\mathrm{Re}\gg 1$ and $L < 1$ this imposes a
strong restriction on the possible lattice sizes.

Respecting the constraints on the parameters we are in the 
position to perform stable simulations. In particular, we can 
resolve a shock on our lattices as Fig.~(\ref{dh:fig1})
demonstrates, which shows a time-slice of a typical configuration. 
We also can observe time dynamics in our simulations as 
Fig.~(\ref{dh:fig3}) shows, and finally, we are able to 
compute the structure functions an example of which is given 
in Fig.~(\ref{dh:fig2}).

\section{IV. Summary and Outlook}
\noindent
We have shown how to perform stable Monte Carlo simulations of
stochastic partial differential equations like the Burgers equation in
the path integral formulation, and in this way demonstrated that it is
possible to calculate observables like the structure functions using an
ensemble--average rather than a time-- or space--average. The lattice
versions of the theories can directly be identified with their continuum
counterparts, and as long as certain constraints on the lattice size are
respected no instabilities occur.
Direct insight into the structures leading to intermittency and, thus,
multiscaling, can be obtained. Especially, we want to point out that
the existence of a dissipation length scale can be observed.

In particular in our simulations of the randomly forced Burgers equation
we observe the following:
\begin{itemize}
\item{Thermalization and autocorrelation times are very long, up to the
order of $10^5$ Monte Carlo steps.}
\item{In the stable runs after thermalization, the typical shock
solutions of Burgers equation and their dynamics can be
observed, see Fig.~(\ref{dh:fig1}) and Fig.~(\ref{dh:fig3}).}
\begin{figure}
\begin{center}
\includegraphics[width=7cm]{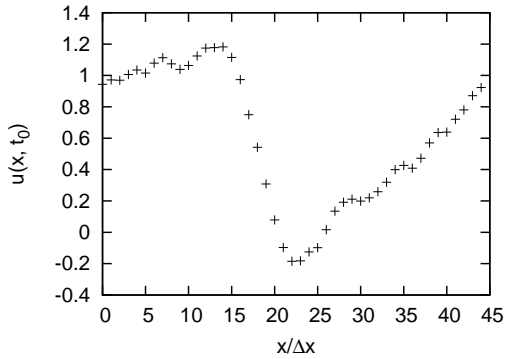}
\caption{Slize of a typical configuration $u(x, t_0)$ at constant
time $t_0$, depending on the (spatial) lattice site $x/\Delta x$;
taken from a $N_{\textrm{x}}=256$, $N_{\textrm{t}}=45$ lattice where
$\mathrm{Re} = 4, \nu = 1/32$.
The typical kink solution associated with the Hopf-eq.~can be clearly 
seen, including a finite shock width due to regularizing viscosity.}
\label{dh:fig1}
\end{center}
\end{figure}\\
\begin{figure}[ht!]
\begin{center}
\vspace{-1.0cm}
\hspace{-1.0cm}
\includegraphics[width=9cm]{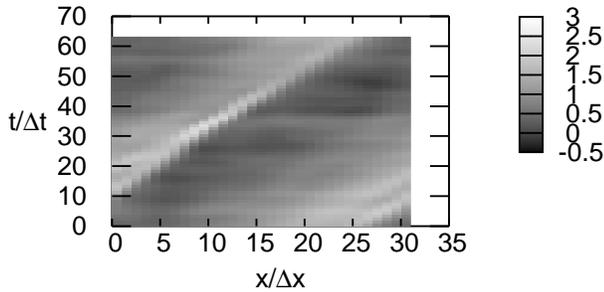}
\vspace{-1.0cm}
\caption{3D-Plot of a configuration $u(x, t)$;
taken from a $N_{\textrm{x}}=32$, $N_{\textrm{t}}=64$ lattice.
The propagation of the shock with constant velocity is clearly visible.}
\label{dh:fig3}
\end{center}
\end{figure}
\item{In the unstable runs before occurence of instabilities,
configurations show typical shock solutions leading eventually to an
energy accumulation near shock positions.}
\item{The distinction between stable and unstable simulations can directly
be related to the existence of a dissipation length scale which is
either bigger (stable) or smaller (unstable
simulations) than the lattice spacing.}
\end{itemize}
First results on the extracted structure functions, e.g.\ see Fig.~(\ref{dh:fig2}) are promising. However, the understanding of systematic errors such as finite size effects or fitting ranges for extracting the scaling exponents from the measured structure functions is still in progress. Nevertheless, we conclude that the path integral
Monte Carlo approach to the Burgers
equation outlined in this article is a feasible and stable tool to
obtain interesting numerical results.
\begin{figure}[!ht]
\begin{center}
\includegraphics[width=7cm]{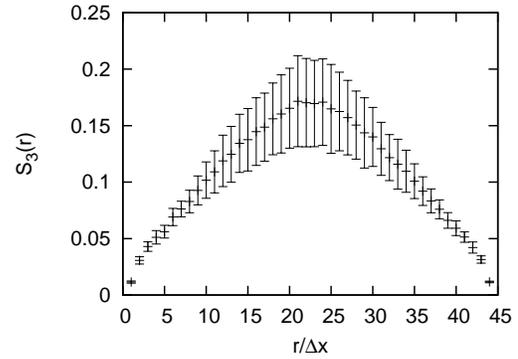}
\caption{Third order structure function $S_3(r)$ as a function of the
separation $r / \Delta x$, evaluated from a simulation with $5*10^6$ configurations
on a $N_{\textrm{x}}=256$, $N_{\textrm{t}}=45$
lattice with $\mathrm{Re} = 16, \nu = 1/512$.}
\label{dh:fig2}
\end{center}
\end{figure}

\acknowledgments
This work is supported by the Deutsche Forschungsgemeinschaft under
project no. Mu 757/14.

\bibliographystyle{eplbib}
\bibliography{burgmc}

\end{document}